# EXPERIMENTAL FAST CHANNEL REACTOR OPERATING IN THE TRAVELING WAVE MODE OF NUCLEAR FISSIONS WITH A SOFT FAST NEUTRON SPECTRUM

V.O. Tarasov*, S.A. Chernezhenko, V.M. Vashchenko, M.R. Shcherbina, V.V. Lavrukhin

*Department of Theoretical and Experimental Nuclear Physics, National Odessa National Polytechnic University, Shevchenko av. 1, Odessa 65044, Ukraine*

___________________________________________________________________________________

**Abstract**

This paper presents the principle scheme of the prototype fast single-channel reactor operating in a traveling wave nuclear fission mode on the soft fast neutron spectrum. The reactor design has cylindrical symmetry. The problem of radiation resistance of construction materials in this design of the experimental reactor is solved on the basis of proposals previously published by the authors of the article. Specifically, the traveling wave of nuclear fission mode is implemented on neutrons with the softened fast spectrum (the spectrum peaks is in range of the 20 – 50 keV), which reduces the neutron impact on construction materials by more than an order of magnitude. To further reduce neutron impact on construction materials, the reactor design incorporates movement of the nuclear fuel relative to the fuel channel walls using elements of the hydravlic fuel handling system. The nuclear fuel in which the fission wave travels is cylindrical uranium dicarbide, which is homogeneous with respect to the neutron field.

## 1 Introduction

A promising direction for the development of nuclear energy is the development of nuclear reactors operating in the traveling wave mode of nuclear fissions. Starting with the work of Feoktistov [1], in which the idea of slow wave burning in a uranium fissile environment was published, all subsequent work in this area until 2015 was devoted to the study of the physical processes of wave burning on fast neutrons. This was due to the fact that in [1] a criterion for wave burning was also formulated and a comparative assessment of the fulfillment of the criterion for thermal and fast neutrons was made, which showed that wave burning of a uranium fissile medium is possible with fast neutrons. In 2013, the American company Terra-Power actively moved into the technical development stage of such a reactor (Traveling Wave Reactor), thereby sparking a scientific debate that identified the problem of radiation resistance of the construction materials of the first wall of the fuel rods, calling into question the possibility of implementing such a nuclear reactor in the near future. Namely, it was shown that the radiation resistance of the construction materials of the first wall of fuel elements of reactors operating in the fast neutron fission traveling wave mode should be ensured at the level of 500 DPA, and the existing reactor construction materials have a radiation resistance of up to 80 DPA and promising materials under development will have it at the level of 100 DPA [2]. To date, two proposals have been made for a possible solution to the problem of radiation resistance of the first wall of fuel rods for the uranium fissile medium. The first proposal was published in 2015 and stated that it was possible to implement a reactor operating in the traveling wave mode of fissions on epithermal neutrons (the maximum energy spectrum of neutrons is in the region of 3–7 eV) [3]. The second proposal is the idea of technical implementation of the movement of the fissile medium, in which the fissions wave runs, relative to the wall of the fuel element at a given speed, ensuring the necessary reduction of the impact of neutron irradiation to a radiation resistance level of 80 - 100 DPA [4].

______________________________
* Corresponding author e-mail: vtarasov@ukr.net

The goal of the work was to develop a basic design for a prototype nuclear reactor, which would make it possible to conduct an experiment for the first time to test the idea of implementing traveling-wave nuclear fission modes in neutron-multiplicating environments, as well as to test the basic technical solutions. Therefore, the authors chose the concept of channel nuclear reactors as a strategy for achieving the goal, which allows for the implementation of a reactor design with one fuel channel at the first stage, and an increase in the number of channels at subsequent stages.

## 2 Experimental fast channel reactor operating in the traveling wave mode of nuclear fissions

Fig. 1 shows the basic design scheme of a prototype fast single-channel reactor operating in the traveling wave mode of nuclear fissions with a soft fast neutron spectrum. The reactor design has cylindrical symmetry (Fig. 1, section of the reactor by the plane in which the axis of symmetry lies). The problem of radiation resistance of construction materials in this design of the experimental reactor is solved using the above-described proposals [3, 4]. More precisely, the mode of a traveling wave of nuclear fissions on neutrons with a softened fast spectrum is implemented (the maximum spectrum is in the range of 20 - 50 keV), which reduces the impact of neutrons on construction materials by more than an order of magnitude, and also to further reduce the impact of neutrons on structural materials the design of the reactor implements the movement of nuclear fuel (Fig. 1, No. 1) relative to the walls of the fuel channel (Fig. 1, No. 4) with the help of construction elements of the hydravlic fuel movement system (Fig. 1, No. 2, No. 25, No. 15 and No. 16) as in [4]. It should be noted that the use of both approaches, each of which, according to estimates (for example, [4]), can solve the problem of radiation resistance, in the opinion of the authors of the article is justified in a prototype reactor to improve the safety of the first experiment.

Fuel cylindrical rod (homogeneous for the neutron field uranium dicarbide in the form of a cylinder with a neutron reflector in a shell made of reactor steel, for example, HT9 or X18H9T) (Fig. 1 No. 1) [5]. As a nuclear fuel in which the fission wave runs the uranium dicarbide, homogeneous with respect to the neutron field, in the form of a cylinder with a diameter of 10 - 20 cm and a length of 200 cm is used. Homogeneous with respect to the neutron field means that the uranium dicarbide contains cylindrical channels for pumping coolant, the diameters of which are much smaller than the diffusion length of neutrons. The diameter of the uranium dicarbide cylinder depends on the characteristics of the neutron reflector, which is included in the design of the fuel rod.

The maximum of the energy soft fast spectrum of neutrons in the region of wave burning of uranium dicarbide is in the region of 20 – 50 keV, which is confirmed by the calculated spectra of neutron moderation in a homogeneous medium of uranium dicarbide [6 – 9].

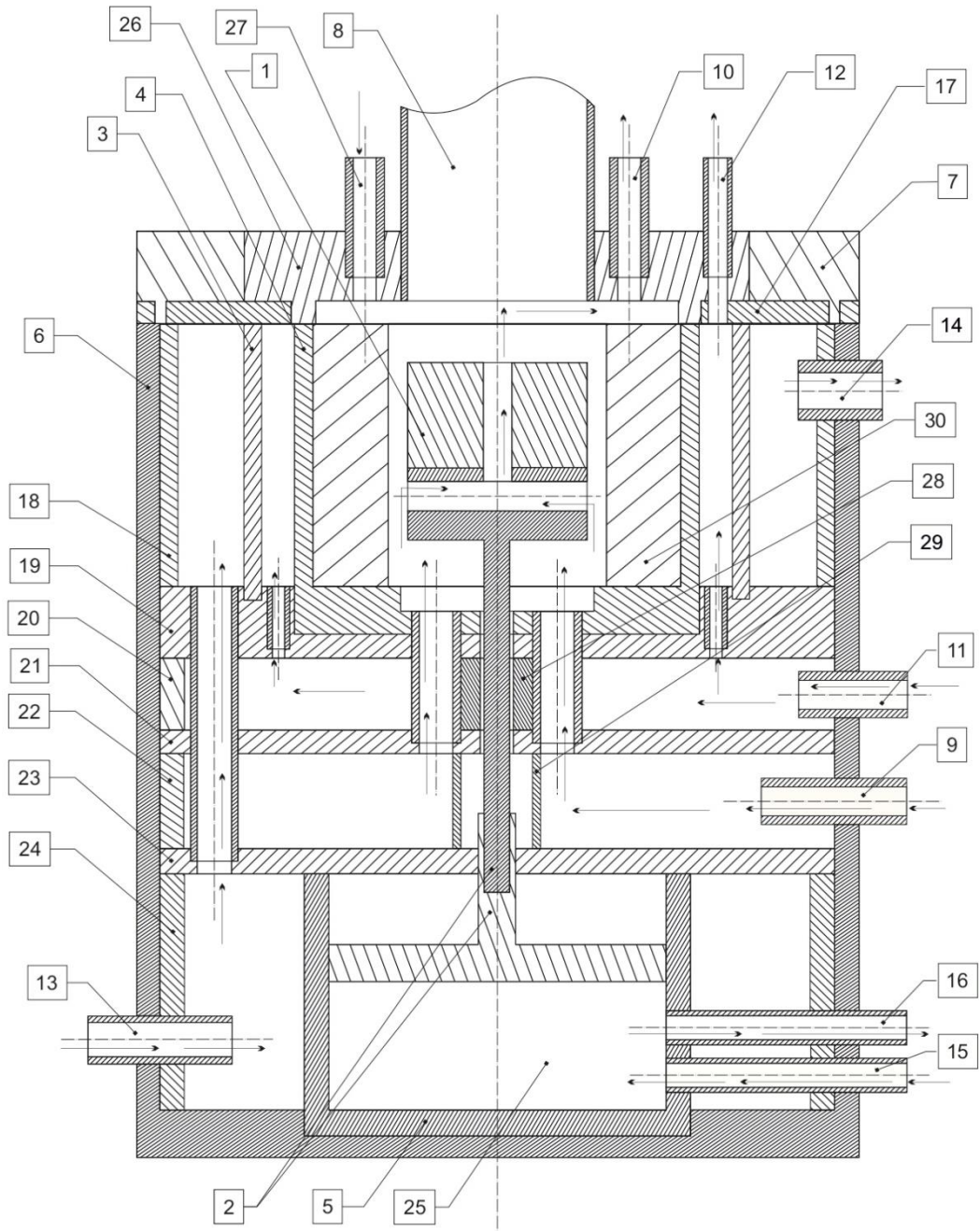

Figure 1. Principled scheme of a prototype reactor with a core with one cylindrical fuel rod, in which the traveling wave mode of nuclear fissions is initiated (1 - cylindrical fuel rod (homogeneous dicarbide in the form of a cylinder with a neutron reflector in a reactor steel shell); 2 - construction of the hydravlic system for moving fuel; 3 - metal wall of fuel channel 2; 4 - metal wall of fuel channel 1; 5 - hydravlic fluid reservoir body; 6 - reactor vessel; 7 - reactor vessel cover; 8 - neutron guide; 9 - inlet pipeline of the internal fuel coolant 1; 10 - outlet pipeline of the internal internal fuel coolant 1; 11 - inlet pipeline of the internal channel heat carrier 2; 12 - outlet pipeline of the internal channel heat carrier 2; 13 - inlet pipeline of the internal reactor vessel heat carrier 3; 14 - outlet pipeline of the internal reactor vessel heat carrier 3; 15 - hydravlic fluid inlet pipeline; 16 - hydravlic fluid outlet pipeline; 17 - 24 - load-bearing cylindrical construction elements of the reactor; 25 - hydravlic fluid; 26 - fuel channel cover; 27 – inlet channel for emergency fuel filling with boron solution to stop the fission chain reaction; 28 - 29 - load-bearing cylindrical construction elements of the reactor; 30 - neutron moderator.

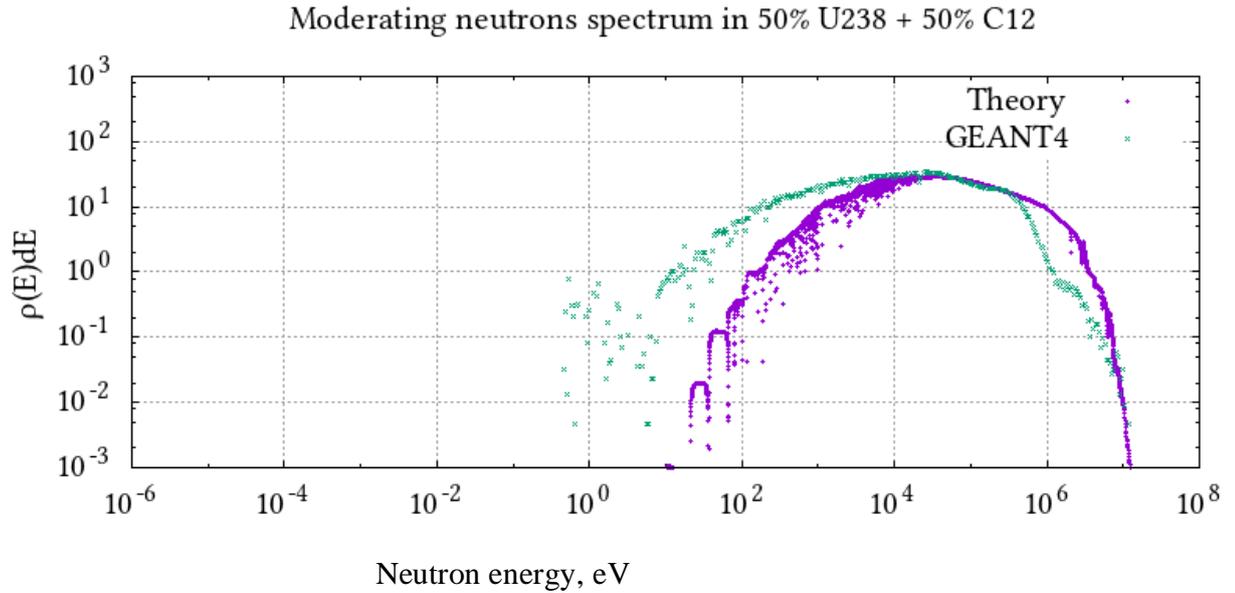

Figure 2. Energy spectrum of neutrons for a uranium-carbon medium (50% uranium 238 and 50% carbon C12), calculated using expression (56) from [8] and using the GEANT4 code [9] with a fission neutron source given by expression (58) from [8], at a moderator temperature of 600 K.

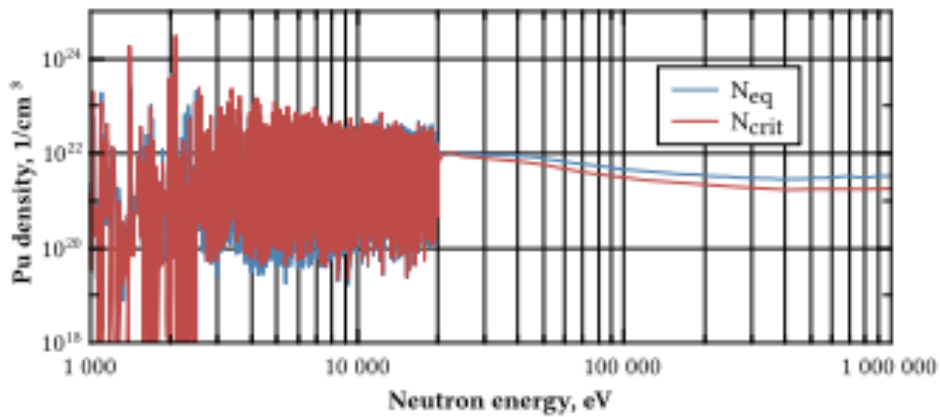

Figure 3. Dependences of the equilibrium and critical concentration of Pu 239 in the neutron energy range from 1 keV to 1 MeV [3].

In works [8, 9] the calculated spectra of neutron moderation in uranium-carbon media of various compositions are presented, and in Fig. 2, as an example, the spectrum of moderating neutrons is presented for an isotropic uniformly distributed source with a fission spectrum in a fissile medium with a composition close to uranium dicarbide.

It should be noted that, according to the results of the study of the fulfillment of the wave burning criterion in uranium fissile media, presented in the work [3] and in Fig. 3 from [3], in the neutron energy range of 20–50 keV (soft fast neutron spectrum), the wave burning criterion is fulfilled ($N_{eq} \geq N_{crit}$), that is, the traveling wave mode of nuclear fissions can be realized in uranium dicarbide.

An estimate of the thermal power for a 10 - 20 cm diameter dicarbide fuel cylinder can be obtained based on the results of mathematical modeling of the traveling wave fission regime presented in [5, 10]. Thus, the authors of the article estimate the thermal power of the developed nuclear reactor prototype at 200 - 1000 MW depending on the diameter of the uranium carbide.

As a neutron source for initiating the traveling wave mode of nuclear fissions in an experiment with a prototype reactor, one can use a subcritical system controlled by a particle accelerator (neutron booster) [11,12] or a compact periodic pulsed nuclear reactor, for example, [13]. For example, in [12] a linear accelerator on a backward wave with proton energy of 1-10 GeV was developed and the length of the accelerator is 28 m.

Figure 4 shows the design scheme of the active zone of the compact pulsed reactor BR-1M in an axial section (1 - upper block, 2 - lower block, 3 - control block, 4 - pulse block, 5-10 protective screens with boron backfill) [13]. The core consists of a highly enriched alloy of uranium-235 with molybdenum (90% enrichment in uranium-235). The neutron flux density is ~ $10^{16} - 10^{20}$ n/cm²•s, the pulse duration is 40 - 100 μs, the core dimensions are: height 26.46 cm and diameter 26.80 cm.

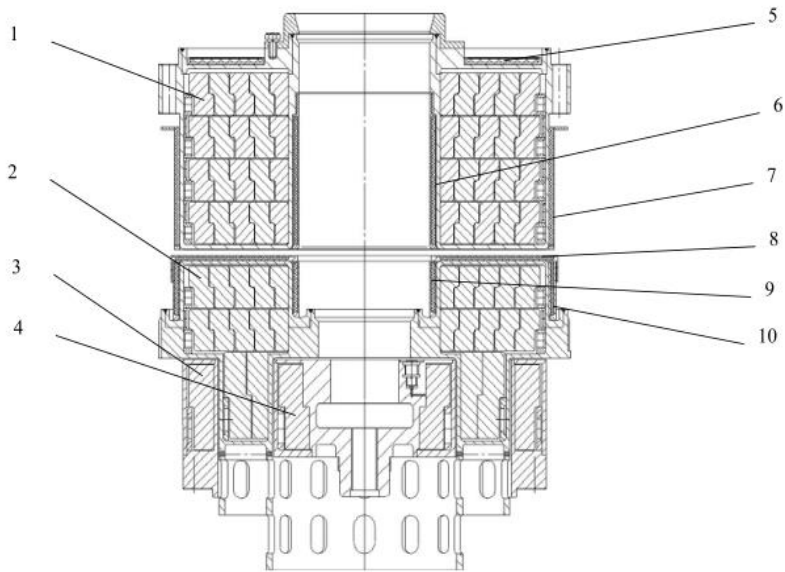

Figure 4. Design scheme of the active zone of the compact pulsed reactor BR-1M in an axial section (1 – upper block, 2 – lower block, 3 – control block, 4 – pulse block, 5-10 protective screens with boron backfill) [13].

Such a neutron source (a compact pulsed nuclear reactor) can be placed on a platform movable above the lid of a multi-channel nuclear reactor, which will allow the initiation of a wave mode of fissions in all channels of the reactor with a single neutron source and independently of each other.

## 3 Basic technical operations for preparing the experimental fast channel reactor for operation

Here is the description of the main operations before preparing the reactor for operation:

1) Using a crane, the cover of the fuel channel No. 26 is removed (Fig. 1), on which the inlet pipes of the neutron guide No. 8, the outlet pipeline of the internal fuel coolant 1 No. 10, the outlet pipeline of the internal channel coolant 2 No. 12 and the inlet channel for emergency fuel filling with boron solution to stop the fission chain reaction No. 13 are located (Fig. 1).

2) Using a lifting crane, fuel assembly No. 1 is loaded into the fuel channel, which is fixed to the structure of the hydravlic fuel movement system No. 2.

3) Using a crane, the fuel channel cover No. 26 is closed.

4) The following coolant pumping systems are included through the reactor: intra-fuel coolant 1, intra-channel coolant 2 and intra-reactor vessel coolant 3, i.e., using pumps through the inlet and outlet channels for coolants 1, 2 and 3, continuous circulation of the reactor coolants is carried out.

5) Using a hydraulic system, we will move the fuel assembly to its highest possible position to initiate the traveling wave mode of neutron-nuclear fissions in the fuel with the help of an external neutron source.

6) Using an external neutron source, the traveling wave mode of neutron-nuclear fissions is initiated in the fuel.

7) After the wave neutron-nuclear burning mode is initiated, i.e. a self-regulating wave of neutron-nuclear fissions running through the fuel is formed, the external neutron source is disconnected or stops its operation and the neutron guide channel branch pipe is closed with a plug, which prevents radiation entering to the working reactor hall.

8) Further, during the designed operating period, until all the fuel is fissioned, using the hydravlic system control system (control of the pump that ensures the supply or pumping of hydraulic fluid to the volume of the hydravlic system), the fuel is moved at a given speed along the reactor fuel channel to its final possible position, which ensures the necessary radiation stability of the wall of channel No. 4, i.e. the reactor operates in a stable operating mode.

9) After the end of the working company, the reactor is stopped, after which work is carried out according to the above-mentioned operations from point 1), i.e., using a lifting crane, the fuel channel cover No. 26 is removed (Fig. 1).

10) Using a lifting crane, fuel assembly No. 1, which is fixed to the structure of the hydravlic fuel movement system No. 2, is unloaded from the fuel channel.

11) Using a crane, the wall pipe of fuel channel No. 4, which has served its working life and requires replacement due to insufficient radiation resistance, is unloaded.

12) A new fuel channel wall pipe No. 4 is loaded using a crane.

13) Using a lifting crane, fuel assembly No. 1, which is fixed to the structure of the hydraulic fuel movement system No. 2, is loaded into the fuel channel.

14) Using a crane, the fuel channel cover No. 26 is closed.

15) The coolant pumping systems through the reactor of intra-fuel coolant 1, intra-channel coolant 2 and intra-reactor vessel coolant 3 is turned on, i.e., with the help of pumps through the inlet and outlet channels for coolants 1, 2 and 3, continuous circulation of the reactor coolants is carried out.

16) Using the hydravlic system, we will move the fuel assembly to its highest possible position to initiate the mode of self-regulating wave of neutron-nuclear fissions in the fuel with the help of an external neutron source.

17) The anti-radiation plug is removed from the neutron guide connection pipe, the neutron guide is connected to the pipe and using an external neutron source, the wave neutron-nuclear burning mode is initiated in the fuel.

18) After the wave neutron-nuclear burning mode is initiated, i.e. a self-regulating wave of neutron-nuclear fissions running through the fuel is formed, the external neutron source is disconnected or stops its operation and the neutron guide channel branch pipe is closed with a plug, which prevents radiation entering to the working reactor hall.

19) Further, during the designed operating period, until all the fuel is fissioned, using the hydravlic system control system (control of the pump that ensures the supply or pumping of hydraulic fluid to the volume of the hydravlic system), the fuel is moved at a given speed along the reactor fuel channel to its final possible position, which ensures the necessary radiation stability of the wall of channel No. 4, i.e. the reactor operates in a stable operating mode.

20) After the end of the working company, the reactor is stopped, after which work is carried out to prepare and start the reactor during the new working company in accordance with the above operations from points 1) - 19) and the reactor operates within the framework of the new working company.

## 4 Conclusions

For the first time the principled scheme of the prototype reactor fast single-channel reactor operating in the traveling wave mode of nuclear fissions with a soft fast neutron spectrum has been developed.

The problem of radiation resistance of construction materials in this design of the experimental reactor is solved as follows: the mode of a traveling wave of nuclear fissions on neutrons with a softened fast spectrum is implemented (the maximum spectrum is in the range of 20 - 50 keV), which reduces the impact of neutrons on construction materials by more than an order of magnitude, and also to further reduce the impact of neutrons on structural materials the design of the reactor implements the movement of nuclear fuel relative to the walls of the fuel channel with the help of construction elements of the hydravlic fuel movement system

The creation of such a prototype reactor would allow for the first time to conduct an experiment to test the idea of implementing traveling-wave nuclear fission modes in neutron-multiplicating environments, as well as to test basic technical solutions.

**References**
[1] Feoktistov L. P. Neutron-fission wave. // Dokl. Akad. Nauk SSSR. – 1989. – Vol. 309, №4. – P. 864 – 867.
[2] Rusov V.D., Tarasov V.A., Sharf I.V. et al, On some fundamental peculiarities of the traveling wave reactor operation. // Science and Technology of Nuclear Installations. – 2015. – Vol. 2015. – P. 1 – 23.
 https://doi: 10.1155/2015/703069
[3] Rusov V.D., Tarasov V.A., Eingorn M.V., Chernezhenko S.A. et al/. Ultraslow wave nuclear burning of uranium-plutonium fissile medium on epithermal neutrons. // Progress in Nuclear Energy.  – 2015. – 83. – P. 105 – 122.
https://doi.org/10.1016/j.pnucene.2015.03.007
[4] V.D. Rusov, V.A. Tarasov, V.N. Vashchenko, S.A. Chernezhenko. Fast traveling-wave reactor of the channel type. // Interdisciplinary Studies of Complex Systems, No. 9 (2016) 36 – 57.
https://doi.org/10.31392/2307-4515/2017-9.3
[5] Shcherbyna M.R. Numerical analysis of self-sustaining traveling wave of nuclear fission propagated by epithermal neutrons in uranium dicarbide medium. / M.R. Shcherbyna, K.O. Shcherbyna, V.O. Tarasov, S.I. Kosenko, S.A. Chernezhenko. // Nuclear physics and atomic energy. – 2024, Vol. 25, № 4, p. 357 – 365.
https://doi.org/10.15407/jnpae2024.04.357
[6]  V. O. Tarasov, S. A. Chernezhenko, A. O. Kakaev et al. On the spectrum of moderated neutrons emitted by the isotropic source in gas fuel. // Nuclear physics and atomic energy. – 2016, Vol. 17, № 4, p. 240 – 249.
https://doi.org/10.15407/jnpae2016.04.240
[7] Rusov V.D., Tarasov V.A., Chernezhenko S.A. et al. The neutron moderation theory, taking into account thermal motions of moderating medium nuclei. // European Physical Journal A - "Hadrons and Nuclei". – 2017, Vol. 53, p. 179 – 192.
DOI 10.1140/epja/i2017-12363-9
[8] Sergey Chernezhenko, Victor Tarasov , Volodymyr Vashchenko et al. Theory of neutron slowing down in reactor media, taking into account inelastic neutron scattering reactions // Research Square, 2025, P. 1–24.
DOI: https://doi.org/10.21203/rs.3.rs-6056230/v1
[9] Agostinelli S. Geant4 - a simulation toolkit. / S. Agostinelli, et al. // Nuclear Instruments and Methods in Physics Research Section A: Accelerators, Spectrometers, Detectors and Associated Equipment. – 2003. – Vol. 506. − Issue 3. − P. 250-303.
https://doi.org/10.1016/S0168-9002(03)01368-8

[10] Viktor Tarasov, Serhiy Chernezhenko, Iryna Korduba, Volodymyr Vashchenko. Simulation of the Traweling Wave Burning Regime on Epithermal Neutrons. // World Journal of Nuclear Science and Technology, 2023, 13, 73-90
https://doi.org/10.4236/wjnst.2023.134006

[11] Taylor Andrew et al. A Route to the Brightest Possible Neutron Source? // Science, Vol. 315, 1092 (2007).
DOI: 10.1126/science.1127185

[12] Bogomolov A.S. et al. Backward wave accelerators as an alternative to classical superconducting accelerators. // Bulletin of Scientific and Technical Development, No. 4 (44), 2011. - P. 1-19.
https://www.xn--h1aellf.xn--p1acf/main/books/BGM.pdf

[13] Arapov A.V., Devyatkin A.A., Drozdov I.Yu., Mochkaev M.V. Results of the physical start-up of the BR-1M reactor. // Problems of high energy density physics. XII Kharitonov Thematic Scientific Readings. Reports. – Sarov: Publishing House of FSUE "RFNC-VNIIEF", 2010, 553 p.
https://www.vniief.ru/vniief/scientific-events/khariton-readings/